\documentclass[aps,prd,preprintnumbers,nofootinbib,twocolumn]{revtex4}
\usepackage{bm}
\usepackage{latexsym}
\usepackage{dcolumn}
\usepackage{amsmath,amsfonts,amssymb}
\usepackage{graphicx,epsfig}
\usepackage{psfrag}
\usepackage{amsthm}

\begin{document}

\title{Rainbow Spacetime from a Nonlocal Gravitational Uncertainty Principle}
\author{Omar El-Refy$^{1,2}$}\email[email: ]{Oelsayed@andrew.cmu.edu}

\author{Syed Masood$^{3}$}\email[email: ]{masood@zju.edu.cn}
\author{Li-Gang Wang$^{3}$}\email[email: ]{lgwang@zju.edu.cn}
\author{Ahmed Farag Ali$^4$}\email[email: ]{ahmed.ali@fsc.bu.edu.eg}

\affiliation{$^{1}$ Department of Physics, Zewail City, Giza, Egypt.\\}
\affiliation{$^{2}$ Department of Physics, Carnegie Mellon University, Pittsburgh, PA 15213, USA.\\}
\affiliation{$^{3}$Department of Physics, Zhejiang University, Hangzhou 310027, China.\\}
 \affiliation{$^{4}$Department  of  Physics,  Faculty  of  Science, Benha  University,  Benha,  13518,  Egypt}


\begin{abstract}
Occurrence of spacetime singularities is one of the peculiar features of Einstein gravity,  signalling limitation on probing short distances in spacetime. This alludes to the existence of a fundamental length scale in nature. On contrary,  Heisenberg quantum uncertainty relation seems to allow for probing arbitrarily small length scales. To reconcile these two conflicting ideas in line with a well known framework of quantum gravity, several modifications of  Heisenberg algebra have been proposed. However, it has been extensively argued that such a minimum length would introduce nonlocality in theories of quantum gravity. In this Letter, we analyze a previously proposed deformation of the Heisenberg algebra (i.e. $p \rightarrow p (1 + \lambda p^{-1})$) for a particle confined in a box subjected to a gravitational field. For the problem in hand, such deformation seems to yield an energy-dependent behavior of spacetime in a way consistent with gravity's rainbow, hence demonstrating a connection between non-locality and gravity's rainbow.
\end{abstract}

\maketitle

It is very well known that a  viable theory of quantum spacetime reconciling principles of quantum physics and Einstein gravity theory is one the foremost goals of modern theoretical physics. Though both of these frameworks work very well in their respective regimes, however, a merger of principles of these two theories results in serious inconsistencies.   One aspect of this contradiction can be seen readily from Heisenberg's uncertainty relation and black hole phenomena. The famous position-momentum uncertainty relation   $\Delta x \Delta p  \geq \hbar / 2$, posits that there exists no lower limit on $\Delta x$. This implies that arbitrarily small distances can be probed using sufficiently high energy scales. However, GR predicts that at sufficiently high energy, a black hole will be formed, which would inevitably prevent the probing process. In contradiction to quantum mechanics, a minimum length (given by the Schwarzschild radius of the black hole to be formed) must exist. Therefore, for any viable theory of quantum gravity reconciling both quantum mechanics and gravity, it must possess a minimum length. As a matter of fact, such a notion of minimum length manifests in almost all candidate theories of quantum gravity. For instance, in loop quantum gravity it is not possible to define an area below a certain minimum\cite{Rovelli}. Also, in string theory, the fundamental string is the smallest possible probe. Hence, it is not possible to probe the geometry of spacetime below the string length scale, which introduces a minimum length to the theory \cite{AMATI,KEMF,MAGG}. In addition, minimal length also manifests in many different approaches to quantum gravity such as Asymptotically Safe Gravity \cite{ROB}, conformally quantized quantum gravity \cite{PAD}, and double field theory \cite{HULL,MAR}.

While such a minimum length is widely believed to be of the order of the Planck length, $l_{p}\approx 10^{-35}m$, it may in fact be several orders of magnitude larger than $l_{p}$ \cite{HOS}. In case this minimum length is much larger than the Planck length, bounds could be imposed on its true value using current experimental data \cite{SURYA1,SURYA2}. In fact, it has been previously suggested that Landau levels and the Lamb shift can be used to obtain bounds for such a length scale \cite{SURYA3}. Also, it has been suggested that an opto-mechanical setup could be used to detect such a minimum length \cite{GUPnature}. The existence of such a minimum length in theories of quantum gravity has many implications, the most important of which is that it introduces nonlocality in the physical theories. In fact, it has been shown to be the case in almost all approaches to quantum gravity, including loop quantum gravity \cite{CAL,MIOT}, perturbative and non-perturbative string theory \cite{CHANG,SANDOR,DOUGLAS,DOD} and the effective field theories based on it \cite{DOD,MOD}.  It is  widely accepted that quantum gravitational effects would break locality at sufficiently high energy scales \cite{RAJU,PAP}. Such nonlocality can be incorporated into quantum gravitational theories via different deformations of the Heisenberg uncertainty principle, and hence the Heisenberg algebra \cite{KEMF}.  A generic form of such an algebra is given by
\begin{eqnarray}
[x^{i}, p_{j}]=i\hbar[\delta_{j}^{i}+f(p)_{j}^{i}]
\end{eqnarray}
where $f(p)_{j}^{i}$ is suitable tensorial function that can be chosen from a specific type of generalized uncertainty principle(variously known as \textit{Gravitational Uncertainty Principle} or \textit{Extended Uncertainty Principle)}. This results in modifications in the coordinate representation of momentum operator. For example, for a quadratic generalized uncertainty principle(GUP), the modified operator representation for momentum $p$ is
\begin{eqnarray}\label{LMBDA}
\tilde{p}\rightarrow i\hbar \partial_{i}(1-\lambda \hbar^2 \partial^{i}\partial_{i})
\end{eqnarray}
with $\lambda$ as the deformation parameter arising from minimal length considerations. In other words, $\tilde{p}$ is the momentum representation at ultrahigh energy scales which reduces to standard quantum representation $p=i\hbar\partial^{i}\partial_{i}$ in the low energy limit when $\lambda\rightarrow 0$. By modifying Heisenberg algebra via this momentum operator representation, we can study the effect of nonlocality using a semi-classical approach, where the gravitational field is treated classically and the matter  is treated quantum-mechanically. Here we use the specific deformation of the Heisenberg algebra where the additional term on right hand side of uncertainty relation  is a properly scaled linear inverse momentum proposed in  \cite{MASOOD}. We demonstrate our result by invoking this modified momentum representation in  the  Schr\"odinger-Newton equation.

Let us consider an astrophysical body  of mass $M$ and radius $R$. Let a probe in the form of a quantum particle (test particle) with energy $E$ and mass $m$ moves on the surface of that astrophysical
body.   The Schr\"odinger-Newton equation for this system can be written as
\begin{equation}
\frac{-\hbar^2}{2m} \frac{d^2\psi}{dx^2} + V(x) \psi(x) = E \psi(x)
\end{equation}
We construct the system in such a configuration that this situation can be approximated as a particle trapped in an infinite potential well,  with gravitational force acting between the walls of that potential well. One can imagine this situation by choosing a particular region of spacetime near that astrophysical body. Further, this spacetime has nonlocal features embodied in it, manifested through the algebraic structure of modified Heisenberg algebra. Thus, for this system, we can write  the potential as
\[
  V(x) =
\begin{cases}
    -kx,&  0\leq x \leq L\\
    0,              & \text{elsewhere}
\end{cases}
\]
Here  $k=\frac{GmM}{R^2}$ and $L$ is  the width of the  infinite potential well.
Now this system can be deformed using the nonlocal deformation of the Heisenberg algebra as \cite{MASOOD}
\begin{eqnarray}
p \rightarrow p(1 + \lambda p^{-1}) \rightarrow (p + \lambda)
\end{eqnarray}
where the parameter $\lambda$ signifies the energy scale or the extent of quantum gravity effects. Using this non-local deformation, the modified   Schr\"odinger-Newton equation (Schr\"odinger equation with gravitational coupling) for this system can be written as
\begin{eqnarray}
\frac{d^2 \psi}{dx^2} + 2 i \alpha \frac{d\psi}{dx} + \beta \Big(x+\frac{E}{k}\Big) \psi = 0
\end{eqnarray}
where $\alpha = \frac{\lambda}{\hbar}$ and $\beta = \frac{2 m k}{\hbar^2}$.
Now we  introduce   $u(x)$, which is  related to $\psi(x)$ by
\begin{eqnarray}
    \psi(x) = e^{-i\alpha x} u(x)
\end{eqnarray}
Thus, we can obtain the following equation for the system,
\begin{eqnarray}
\frac{d^2 u(x)}{dx^2} + \beta \Big(x + \frac{E}{k}\Big) u(x) = 0
\end{eqnarray}
 Furthermore,  we let $s = - \beta^{\frac{1}{3}} \Big(x + \frac{E}{k}\Big)$
and obtain
\begin{eqnarray}
    \frac{d^2 u(s)}{ds^2} - s u(s) = 0
\end{eqnarray}
Thus, we can write the solution to this equation as
\begin{eqnarray}
    u(s) = C_1 Ai(s) + C_2 Bi(s)
\end{eqnarray}
where $Ai(s)$ and $Bi(s)$ are the Airy functions of first and second kinds, respectively, and $C_1$ and $C_2$ are two constants.
Now we obtain the following solution for $\psi(x)$,
\begin{eqnarray}\nonumber
    \psi(x) = e^{-i \alpha x} \Big\{C_1 Ai\Big[- \beta^{\frac{1}{3}} \Big(x + \frac{E}{k}\Big)\Big] \\
    + C_2 Bi\Big[- \beta^{\frac{1}{3}} \Big(x + \frac{E}{k}\Big)\Big] \Big\}
 \end{eqnarray}
Using the boundary conditions,    $\psi(0) = 0$ and $\psi(L) = 0$ (as it is approximated by an infinite potential well, with gravitational potential  inside it),
we obtain  from above
\begin{eqnarray}\nonumber
    \psi(x) = C e^{-i \alpha x} \Big\{Bi\Big[-\frac{\beta^{\frac{1}{3}} E}{k}\Big] Ai\Big[- \beta^{\frac{1}{3}} \Big(x + \frac{E}{k}\Big)\Big]\\
    \nonumber- Ai\Big[-\frac{\beta^{\frac{1}{3}} E}{k}\Big] Bi\Big[- \beta^{\frac{1}{3}} \Big(x + \frac{E}{k}\Big)\Big] \Big\}
\end{eqnarray}
where $C = \frac{C_1}{Bi\Big[-\frac{\beta^{\frac{1}{3}} E}{k}\Big]}$.

We now consider the limiting case, where $E \rightarrow \infty$ (i.e. for extremely high energies), which would imply that the arguments
of the Airy functions would approach $-\infty$. In this limit, the two functions $Ai(x)$ and $Bi(x)$ exhibit a sinusoidal behavior, namely
\begin{eqnarray}
    Ai(x) && \sim \frac{1}{\sqrt{2 \pi} x^{\frac{1}{4}}} \sin\Big(\frac{2}{3} |x|^{\frac{3}{2}} + \frac{\pi}{4}\Big) \\
     Bi(x) && \sim \frac{1}{\sqrt{2 \pi} x^{\frac{1}{4}}} \cos\Big(\frac{2}{3} |x|^{\frac{3}{2}} + \frac{\pi}{4}\Big)
\end{eqnarray}
as $x \rightarrow -\infty$. Using  this asymptotic behavior, we obtain
\begin{center}
\begin{eqnarray}
    \psi(x) & \sim C e^{-i \alpha x} \frac{1}{2\pi} \Bigg[\frac{k}{\beta^2 E \Big(x+\frac{E}{k}\Big)}\Bigg]^{\frac{1}{4}} \nonumber \\
    & \times \Bigg[\cos\Bigg\{\frac{2}{3} \Big|\frac{\beta E}{k}\Big|^{\frac{3}{2}} + \frac{\pi}{4}\Bigg\} \sin\Bigg\{\frac{2}{3} \Big|\beta(x+\frac{E}{k})\Big|^{\frac{3}{2}} + \frac{\pi}{4} \Bigg\} \nonumber \\
  & - \sin \Bigg\{\frac{2}{3} \Big|\frac{\beta E}{k}\Big|^{\frac{3}{2}} + \frac{\pi}{4}\Bigg\}  \cos \Bigg\{\frac{2}{3} \Big|\beta(x+\frac{E}{k})\Big|^{\frac{3}{2}} + \frac{\pi}{4} \Bigg\}  \Bigg]
\end{eqnarray}
\end{center}

Now, applying the second boundary condition (i.e. $\psi(L) = 0$) to the above asymptotic expression and rearranging, we obtain
\begin{eqnarray}\nonumber
   & \cos\Big\{\frac{2}{3} \Big|\frac{\beta E}{k}\Big|^{\frac{3}{2}} + \frac{\pi}{4} \Big\} \sin\Big\{\frac{2}{3} \Big|\beta L + \frac{\beta E}{k}\Big|^{\frac{3}{2}} + \frac{\pi}{4} \Big\} \\
   & = \sin \Big\{\frac{2}{3} \Big|\frac{\beta E}{k}\Big|^{\frac{3}{2}} + \frac{\pi}{4} \Big\} \cos\Big\{\frac{2}{3} \Big|\beta L
   + \frac{\beta E}{k}\Big|^{\frac{3}{2}} + \frac{\pi}{4} \Big\}
\end{eqnarray}
in the limit $E \rightarrow \infty$.  Now, since the above equation holds for all $\beta$, $E$, $k$ and $L$ that lie in the asymptotic region
we assume, we must have
\begin{equation}
    \frac{2}{3} \Big|\frac{\beta E}{k}\Big|^{\frac{3}{2}} + \frac{\pi}{4} = \frac{2}{3} \Big|\frac{\beta E}{k} + \beta L\Big|^{\frac{3}{2}} + \frac{\pi}{4} + 2 \pi n
\end{equation}
where $n = 0, \pm 1 , \pm 2 , \pm 3, ...$ Now, noting that $\beta$, $E$, $k$ and $L$ are all positive,  so we can express  $L$ as
\begin{eqnarray} \label{box-length}
    L = \Big[\Big(\frac{E}{k}\Big)^{\frac{3}{2}} + \frac{3 \pi n}{\beta^{\frac{3}{2}}}\Big]^{\frac{2}{3}} - \frac{E}{k}
\end{eqnarray}
where $n \in \mathcal{Z}$, an integer.
Next, we note that, for $L$ to be of positive-definite value, one must have
\begin{eqnarray}
    \Big[\Big(\frac{E}{k}\Big)^{\frac{3}{2}} + \frac{3 \pi n}{\beta^{\frac{3}{2}}}\Big]^{\frac{2}{3}} > \frac{E}{k}
\end{eqnarray}
which upon using $\beta = \frac{2 m k}{\hbar^2}$ and $\hbar = 1$, simplifies to
\begin{eqnarray} \label{condition}
    \Big(1 + \frac{3 \pi n}{(2 m)^{\frac{3}{2}} E^{\frac{3}{2}}}\Big)^{\frac{2}{3}} > 1
\end{eqnarray}
Obviously, (\ref{condition}) is not satisfied for zero and negative values of $n$, which rules them out as non-physical states. Thus, only $n = 1, 2, 3, ...$ are allowed.
Now, since we are adressing ultra-high energy limit, (\ref{box-length}) could take a much simpler form. To elucidate, using $\beta = \frac{2 m k}{\hbar^2}$ and $\hbar = 1$, we re-write it as
\begin{eqnarray}
            L = \frac{E}{k} \Big(1 + \frac{3 \pi n}{(2 m)^{\frac{3}{2}} E^{\frac{3}{2}}}\Big)^{\frac{2}{3}} - \frac{E}{k}
\end{eqnarray}
Obviously, the second term inside the brackets is extremely small, which enables us to Taylor-expand the whole bracket up to first order, which, upon inserting the value of $k$ further simplifies to
\begin{eqnarray} \label{final}
   L = \frac{n\pi R^2}{G M \sqrt{2m^5E}}
\end{eqnarray}
Above relation indicates that box length is quantized in an energy-dependent way, whilst  gravity constant $G$ enters this quantization scheme. It puts a limitation on the way we find the particle in the box. For a box length different than what above relation implies, there is no meaning to the existence of a particle. It must be emphasized here that box length is purely a geometrical aspect and the box under consideration here is a hypothetical region of spacetime. If this length is explicitly energy dependent as seen in the above relation, then geometry of spacetime as seen by the probe depends on its energy.

In view of the above result, it can be shown that that the value of Newton's constant becomes dependent on the energy of the spacetime probe through the relation  (\ref{final}). Hence, any metric geometry shall pick up the energy dependence. This energy dependent Newton's constant signifies the conjecture that the effective gravitational coupling might depend on the energy scale and satisfy a condition of renormalization group flow. This is what is essentially entailed by the theory of rainbow gravity. To put it simply, a particle of energy $E$ propagating in such a spacetime sees a geometry that depends upon its own energy $E$.  This results in a spacetime geometry characterized by the metric\cite{SMOLIN}
\begin{eqnarray}
g_{\mu\nu}(E)=\eta^{\mu\nu}e_{\mu}(E) \otimes e_{\nu}(E)
\end{eqnarray}
where $e_{\mu}$ and $e_{\nu}$ are orthonormal frame fields and $E$ is the energy of the probing particle. In fact, such a behavior of spacetime geometry can be motivated from the the extension of Doubly Special Relativity\cite{CAMELIA33} on curved spacetime background, where the energy-momentum dispersion relation modifies as
\begin{eqnarray}
E^2 f^2(l_{pl}E)-   p^2g^2(l_{pl}E)=m_{0}^2
 \end{eqnarray}
 Here $f(l_{pl}E)$ and $g(l_{pl}E)$ are called rainbow functions that depend on the energy of the probing particle with $l_{pl}$ as Planck length. In this formulation, Einstein field equations of GR (in $c=1$ units) read as
 \begin{eqnarray}
 G_{\mu\nu}(E) - g_{\mu\nu} \Lambda(E)=8 \pi G(E) T_{\mu\nu}
 \end{eqnarray}
 Note the energy dependence of Newton's constant, $G(E)$ and cosmological constant, $\Lambda(E)$.
 Such effects are much more pronounced near the quantum gravity scale, for $\lambda$ in (\ref{LMBDA}) is a deformation parameter, which essentially deforms the theory significantly near the this energy scale, while preserving the familiar classical limit of GR at low energies, implying both $f(l_{pl}E)\rightarrow 1$ and $g(l_{pl}E)\rightarrow 1)$ at low energies $E$. However, as the minimal measurable length in string theory can be several orders of magnitude above Planck scale, $l_{pl} = g_s^{1/4} l_s$ \cite{HOS}, this energy can also be several orders of magnitude below Planck energy (as here $\beta$ would also be several magnitudes above Planck scale). This energy could be bounded by astrophysical observations, and it would be interesting to analyze such observations, and obtain certain bounds on the parameters that signify energy dependence of the metric.

Meanwhile, to appreciate the impact of nonlocality on particle dynamics and spacetime geometry, wavefunctions for different values of $n$ and $\lambda$ are plotted in
Fig.1 and Fig.2. As evident from Fig.2, the deformation parameter, $\lambda$, controls the allowable box lengths (acting as the probing energy scale indicator). However, in the scenario of ultra-high energy regime, energy dependence of box length $L$ is absorbed into particle energy $E$ as depicted in  (\ref{final}). Though the form of sinusoidal nature of box wavefunctions remain unaffected compared to the conventional box physics, as can be seen from Fig.1, we however observe the fall of amplitude for particle wavefunctions  as the parameter $\lambda$ increases. This is tantamount to saying that particle  becomes more and more confined in a particular region of spacetime as the parameter $\lambda$ increases. However, beyond a certain value of $\lambda$, wavefunction might lose its meaning, given the occurrence of spacetime singularities or the existence of a minimal measurable length scale. The above result is potentially alluding to the breakdown of our present theories to explain the nature of spacetime at the quantum gravity scale.
\begin{figure}[t!]
\includegraphics[width = 0.8 \linewidth]{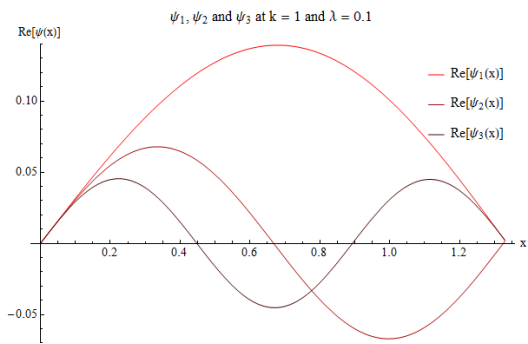}
 \caption{Real part of ground, first and second excited states for $k=1$ and $\lambda=0.1$}
\label{chapt71}
\end{figure}

\begin{figure}[t!]
\includegraphics[width = 0.73 \linewidth]{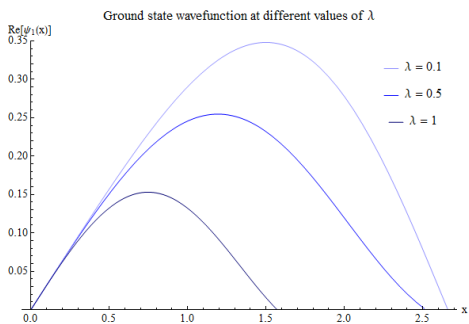}
 \caption{Plot of real part of ground state wavefunction for $\lambda=0.1,0.5$ and $1$, where $k = 1$ for all cases}
\label{chapt71}
\end{figure}

 Several quantum gravity approaches can be taken to justify gravity's rainbow emergence from nonlocality, such as string theory. It is very known that due to renormalization group flow in a quantum field theory, the coupling constants flow and thus depend on the scale at which the theory is probed \cite{ROST,WARNER}. However, the scale at which a theory is probed depends on the energy of the probe. Therefore, the coupling constants depend explicitly on the scale at which the theory is probed and implicitly on the energy of the probe used to probe the theory. Since string theory can be analyzed as a two-dimensional conformal field theory, where the target space metric is regarded as a matrix of coupling constants of the theory, coupling constants should in turn depend on the scale at which the theory is being probed. And since the scale at which the theory is being probed is equivalent to the energy of the probe, it can be argued from string theory that the geometry of spacetime should depend on the energy of the probe \cite{SMOLIN,FARAG1}.
Arguably, such energy-dependent spacetime deformation should be manifested in various other approaches to quantum gravity. For instance, gravity's rainbow has been motivated from the results obtained in loop quantum gravity and $\kappa$-deformed Minkowski spacetime \cite{AMELINO1,AMELINO2}. In addition, deformations of the energy-momentum dispersion relation appear in the Horava-Lifshitz gravity \cite{HORAVA1,HORAVA2}, discrete spacetime \cite{Hooft}, models based on string field theory \cite{KOSTEL}, spacetime foam \cite{Amelino3}, spin-network \cite{RODO}, non-commutative geometry \cite{CAROLL,FAIZAL1}, and ghost condensation \cite{FAIZAL2}. It may also be noted that Greisen-Zatsepin-Kuzmin limit (GZK limit), with which the Pierre Auger Collaboration and the High Resolution Fly's Eye (HiRes) experiment are consistent \cite{ABRA}, has been used to argue for such a deformation of the energy-momentum dispersion relation \cite{KG,ZAT}. In fact, several different tests have been proposed to experimentally verify this idea \cite{Ali555}. In addition, an explanation of the hard spectra of gamma ray bursts has been previously proposed \cite{Amelino3}. Hence, there are strong motivations of both gravity's rainbow and non-locality from numerous approaches to quantum gravity. In this work, we have briefly demonstrated one such way by which gravity's rainbow could emerge from the nonlocality in spacetime.

\begin{acknowledgements}
LGW would like to acknowledge support from Zhejiang Provincial Natural Science Foundation of China under Grant No. LD18A040001
and National Natural Science Foundation of China under Grant No's. 11674284 and 11974309.
\end{acknowledgements}

\end{document}